\documentstyle[prb,aps,twocolumn,psfig]{revtex}

\begin{document}
\wideabs{
\title{ The two-dimensional quantum Heisenberg antiferromagnet:
	\\ effective Hamiltonian approach to the thermodynamics }

\author{Alessandro Cuccoli\cite{e-AC} and Valerio Tognetti\cite{e-VT}}
\address{Dipartimento di Fisica dell'Universit\`a di Firenze
	and Istituto Nazionale di Fisica della Materia (INFM),
	\\ Largo E. Fermi~2, I-50125 Firenze, Italy}

\author{Ruggero Vaia\cite{e-RV}}
\address{Istituto di Elettronica Quantistica
	del Consiglio Nazionale delle Ricerche,
	via Panciatichi~56/30, I-50127 Firenze, Italy,\\
	and Istituto Nazionale di Fisica della Materia (INFM).}

\author{Paola Verrucchi\cite{pres-addr}}
\address{Istituto di Elettronica Quantistica
	del Consiglio Nazionale delle Ricerche,
	via Panciatichi~56/30, I-50127 Firenze, Italy,\\
	and ISIS Facility, Rutherford Appleton Laboratory,
	Chilton, Didcot, Oxfordshire OX11 0QX, U.K.\\ \phantom{x}}

\date{October 23, 1997}

\maketitle

\begin{abstract}
In this paper we present an extensive study of the thermodynamic properties
of the two-dimensional quantum Heisenberg antiferromagnet on the square
lattice; the problem is tackled by the pure-quantum self-consistent
harmonic approximation, previously applied to quantum spin
systems with easy-plane anisotropies, modeled to fit the
peculiar features of an isotropic system.
Internal energy, specific heat, correlation functions, staggered
susceptibility, and correlation length are shown for different values of the
spin, and compared with the available high-temperature expansion and
quantum Monte Carlo results, as well as with the available experimental
data.
\end{abstract}


} 

\section{Introduction}
\label{s.intro}

The fully isotropic Heisenberg model may well be considered the cornerstone
of modern theory of magnetic systems; the reason for such an important role
is the simple structure of this model's Hamiltonian, whose high symmetry is
responsible for most of its peculiar features.
Recent years have seen a growing interest in the specific case of the
two-dimensional quantum Heisenberg antiferromagnet (2DQHAF) on the square
lattice, due both to its theoretically challenging properties and to its
being the best candidate for modelling the magnetic behaviour of the parent
compounds of some high-$T_{\rm{c}}$
superconductors\cite{SokolP93,Chakravarty90}.

As for the theory, the 2DQHAF cannot exhibit long range order (LRO) for
$T>0$ because of its being a two-dimensional model with a continuous
symmetry (Mermin and Wagner theorem\cite{MerminW66}); the study of the
finite temperature paramagnetic phase is hence a matter of understanding
what kind of disorder one is dealing with, i.e., what kind of correlation
exists amongst magnetic moments on different sites. At $T=0$ quantum
fluctuations make the system change from the classical-like N\'eel state to
a ground state that can be rigorously proven\cite{NevesP86} to be
ordered for $S{\ge}1$; for $S=1/2$ the situation is not clear yet,
although more or less direct evidences for an ordered ground state, even in
this extreme quantum case, can be drawn from many different studies (for a
review, see, for instance, Ref.~\onlinecite{Manousakis91}), including the
present one (see also Ref.~\onlinecite{CTVV96prl}).

The experimental activity stems from the existence of several real
compounds whose crystal structure is such that the magnetic ions form
parallel planes and interact strongly only if belonging to the same plane.
As a consequence of such structure, their magnetic behaviour is indeed
two-dimensional down to those low temperatures where the weak inter-plane
interaction becomes relevant, driving the system towards a
three-dimensional ordered phase; an antiferromagnetic Heisenberg
interaction and a small spin value make these compounds 2DQHAF's.
This is indeed the case of high-$T_{\rm{c}}$ superconductors of the
La$_2$CuO$_4$ or Sr$_2$CuO$_2$Cl$_2$ family ($S=1/2$), of other magnets
such as the $S=1$ La$_2$NiO$_4$ and K$_2$NiF$_4$, and the $S=5/2$
Rb$_2$MnF$_4$. The interplane interaction in these compounds is several
orders of magnitude smaller than the intra-plane one, thus offering a large
temperature region where the two-dimensional behaviour can be safely
studied. Furthermore, their having different values of the spin allows a
meaningful analysis of the spin dependence of the thermodynamic properties,
which is essential if the interplay between thermal and quantum
fluctuations is to be clarified.

In order to go beyond the very first treatments (mainly, mean field theory
and spin wave theory), that are mostly not satisfactory in coping with its
strong nonlinearity, the QHAF has been tackled by the field theory of
the quantum nonlinear $\sigma$ model (QNL$\sigma$M)
\cite{Haldane83Affleck85,ChakravartyHN89,HasenfratzN91}. Nevertheless, the
approximations needed to reduce the actual spin model to the related field
theory are drastic, as they usually approximate the essential features of
magnetic systems in solid state physics, i.e., discreteness, strong
nonlinearity of the Hamiltonian, and appearance of the angular momentum
operators (that is, in a classical-like language, sphericity of the phase
space). This inadequacy becomes more evident when experimental data are
available and a quantitative comparison with theoretical predictions is
attempted; realistic spin models and more refined methods must then be
used, and the number of fit parameters minimized to let the real compounds
drive the overall comprehension of the problem. Indeed, despite the success
in explaining some of the early experimental data, the QNL$\sigma$M
approach does not lead to a satisfactory understanding of the problem when
higher values of the spin and higher temperatures are to be
considered\cite{Elstner97Etal95}.

This work follows a previous paper \cite{CTVV96prl}, and, together with
the latter, it is our attempt to move forward studying the finite
temperature properties of the 2DQHAF by the effective Hamiltonian method
based on the pure-quantum self-consistent harmonic approximation
(PQSCHA) as developed in Ref.~\onlinecite{CTVV92ham}; we report here on the
detailed derivation of the effective spin Hamiltonian and more results are
shown as temperature and spin value are varied, together with experimental
data\cite{GrevenEtal94,NakajimaEtal95,CarrettaRS97,BirgeneauEtalunpub97},
classical\cite{ShenkerT80,Kim94} and quantum Monte Carlo
\cite{MakivicD91,KimLT97} (MC) simulations and high-temperature
expansion\cite{Elstner97Etal95} (HTE) results. No best-fit procedure is
involved in the comparison with the experimental data, as we just need to
know the spin model, which is unambiguously defined, in the case of the
QHAF on the square lattice, once the values of the spin and of the exchange
integral are given. The agreement we find with the available data, together
with the clarity of our approach, allows us to draw a comprehensive picture
of the subject, including both the analysis of the 2DQHAF behaviour and the
discussion of previous approaches used for the same purpose.

In Sec.~\ref{s.qha} we introduce the 2DQHAF model and briefly describe
the problems encountered by alternative approaches.
In Sec.~\ref{s.eh} the effective Hamiltonian method is described, and
the approximations involved in this specific implementation are discussed;
the effects of quantum fluctuations on the physics of the 2DQHAF are then
analysed in terms of the quantum renormalizations introduced by the PQSCHA.
Sections~\ref{s.tp} and~\ref{s.expt} contain our results for the
thermodynamic properties (internal energy, specific heat, correlation
functions, correlation length, and staggered susceptibility) compared with
MC and HTE results, as well as with the experimental data. Conclusions are
drawn in Sec.~\ref{s.concl}.

\section{The quantum Heisenberg antiferromagnet}
\label{s.qha}

The two-dimensional quantum Heisenberg Antiferromagnet is described by the
Hamiltonian
\begin{equation}
 \hat{\cal H}= {J\over2} \sum_{{\bf{i}},{\bf{d}}}
 \hat {\mbox{\boldmath$S$}}_{\bf{i}}{\cdot}
 \hat {\mbox{\boldmath$S$}}_{{\bf{i}}+{\bf{d}}}~,
\label{e.ham}
\end{equation}
where $J$ is positive and the quantum spin operators
$\hat{\mbox{\boldmath$S$}}_{\bf{i}}$ satisfy
$|\hat{\mbox{\boldmath$S$}}_{\bf{i}}|^2=S(S{+}1)$.
The index ${\bf{i}}\equiv(i_1,i_2)$ runs over the sites of a square
lattice, and ${\bf{d}}$ represents the displacements of the 4
nearest neighbours of each site, $(\pm{1},0)$ and $(0,\pm{1})$.

The most important feature of this model is the O(3) symmetry of its
Hamiltonian, implying no spontaneously broken symmetry for $T>0$: the
system does not support LRO at finite temperature and the standard
spin-wave theory consequently produces unphysical results.
The existence of local alignment directions, due to the persistence of
strong short-range order up to high temperatures, makes possible
the definition of a properly modified spin-wave theory\cite{Takahashi89}
whose results are remarkably good. However, spin-wave theory remains quite
an inadequate tool to study the thermodynamics of the 2DQHAF and this
indeed stimulated several authors to search for alternative theories.

It is useful to see how disorder develops in the model described by
Eq.~(\ref{e.ham}) when temperature is switched on:
the ground state, hereafter {\it assumed ordered} for any spin value,
becomes unstable because of long-wavelength excitations that
gradually flip the spins as one moves far from the chosen origin of the
lattice; their energy is small and they do not disturb the {\it local}
order of the system. They have a classical character in that their
contribution to the thermodynamics is almost the same in both the classical
and the quantum case. What makes the latter different is the additional
local disorder introduced by the short-range purely quantum fluctuations,
whose renormalizing effect decreases as temperature increases.

The 2DQHAF model (\ref{e.ham}) has been studied by several authors in terms
of the QNL$\sigma$M field theory, involving a three-component vector field
${\mbox{\boldmath$\Omega$}}$ with the constraint $|{\mbox {\boldmath
$\Omega$}}|=1$ and spatial integrations subjected to a short-distance
cutoff $\Lambda^{-1}$. The model depends on two parameters: the bare {\it
spin stiffness} $\rho_S^0$ and the bare {\it spin-wave velocity} $c_0$; the
coupling constant turns out to be $g_0=c_0\Lambda/\rho^0_S$. Despite their
names, the parameters $\rho_S^0$ and $c_0$ are {\it not} directly related
with those ($J$ and $S$) defining the 2DQHAF: this relation is indeed the
weakest point of the QNL$\sigma$M approach.

A first link between the two models was established by Haldane and Affleck
\cite{Haldane83Affleck85} under a large-$S$ condition, i.e., in the
semiclassical limit. Their mapping gives $\Lambda$, $\rho_S$ and $c_0$ as
$a^{-1}$, $JS^2$, and $2\sqrt{2}JSa$, respectively ($a$ is the lattice
spacing); it follows that $g_0=2\sqrt{2}S^{-1}$, which means that the
semiclassical limit of the spin model ($S{\gg}1$) corresponds to the weak
coupling regime of the field theory ($g_0{\ll}1$). The validity of this
approach when studying real compounds is more than questionable, and
Haldane's suggestion of replacing $S$ by $\sqrt{S(S+1)}$ does not solve the
problem.

The way Chakravarty, Halperin, and Nelson (CHN)\cite{ChakravartyHN89}
connected the two models has greater generality. They used symmetry
arguments to show that the long-wavelength physics of the QHAF must be the
same of that of the QNL$\sigma$M, a result that holds regardless of the spin
value. However, they could not define the field theory parameters in terms
of those of the spin system; therefore the spin stiffness and spin wave
velocity are just phenomenological fitting parameters to be determined from
either experiments or simulations.

The analysis carried out by CHN on the QNL$\sigma$M leads to the
characterization of three different regimes, called quantum
disordered, quantum critical (QCR) and renormalized classical
(RCR), the most striking difference amongst them being the temperature
dependence of the spin correlations. If $g_0$ is such as to guarantee LRO
at $T=0$, the QNL$\sigma$M is in the RCR and its long-wavelength (i.e.,
low-temperature) physics is that of the classical model with parameters
renormalized by quantum fluctuations
($\rho^0_{\rm{S}}{\rightarrow}\rho_{\rm{S}}$, $c_0{\rightarrow}c$) and a
short-wavelength cutoff of order $\alpha\equiv{c}/T$. As far as the
correlation length $\xi$ is concerned, the famous low-temperature two-loop
result for the renormalized classical regime is
\begin{equation}
 \xi_{\scriptscriptstyle\rm{CHN}}=C_\xi\left({c\over2\pi\rho_S}\right)
 \exp (2\pi\rho_S/T)~,
\label{e.xiCHN}
\end{equation}
where $C_\xi$ is a nonuniversal coefficient.
Hasenfratz and Niedermayer\cite{HasenfratzN91} have subsequently
calculated the leading correction in $T/2\pi\rho_S$ to Eq.~(\ref{e.xiCHN}),
and also found the coefficient $C_\xi=ea/8$, so that
\begin{equation}
 \xi_{\scriptscriptstyle\rm{HN}}= \xi_{\scriptscriptstyle\rm{CHN}}
 \left[1-{T\over4\pi\rho_S}+O\left({T\over2\pi\rho_S}\right)^2\right];
\label{e.xiHN}
\end{equation}
although Eq.~(\ref{e.xiHN}) extends the temperature region where
Eq.~(\ref{e.xiCHN}) can be used, both of them do not work for intermediate
or high temperature. The fact that the
low-temperature experimental results for $S=1/2$ two-dimensional magnetic
compounds can be fitted by those of a 2DQNL$\sigma$M in the renormalized
classical regime, has lead to assert that the ground state of the
2DQHAF is ordered also in the $S=1/2$ case.

The results by CHN are of great general value and have been extensively
used to understand the $S=1/2$ experimental data, but their approach has
several substantial drawbacks for that purpose. First of all, for precisely
given $S$ and $J$ of the real compound, the fundamental parameters $\rho_S$
and $c$ remain unknown; the necessary best fit procedure involved in their
determination introduces a substantial uncertainty in the whole of the
work. Secondly, the restriction to low temperatures cannot be avoided, thus
making the HTE technique, extensively used by Elstner {\it
et~al.}\cite{Elstner97Etal95}, of great help and importance in this
framework. Finally, things get worse for higher values of the spin, as has
been recently pointed out by Elstner {\it et~al.}\cite{Elstner97Etal95};
further adjustments of the fit parameters are necessary to reproduce the
experimental data\cite{GrevenEtal94} and the dependence of $\xi$ upon $S$
cannot be analyzed because it is not directly addressed by the theory.
Quantum MC simulations\cite{MakivicD91,KimLT97} provide quite good results
in a rather large temperature region, but up to now they are only available
for $S=1/2$.

As far as the different regimes of the 2DNL$\sigma$M are concerned, the RCR
is the most interesting (being the one linked with the real $S{\ge}1/2$
compounds), but the QCR has also attracted much interest in recent years.
CHN found that any 2DQNL$\sigma$M with an ordered ground state crosses over
from the RCR to the QCR at sufficiently high temperature. This statement
cannot be trivially extended to the 2DQHAF, as we know that the relation
between the two models only holds for low temperature. Nevertheless, the
lower the spin, the lower the temperature at which such a crossover should
occur, so that for sufficiently small $S$ at least some signs of an
intervening QCR-like regime could be detected\cite{ChubukovSY94}. As the
correlation length of the 2DQNL$\sigma$M in the QCR is
$\xi{\propto}\alpha(T)=c/T$, such a sign could be, for instance, a
temperature dependence of the measured $\xi$ which becomes less pronounced
as $T$ increases. No such experimental evidence exists for pure compounds,
but some doped materials show a similar behaviour\cite{KeimerEtal92}; it
has been argued that, by thinking of the doping as causing an effective
increase of the quantum coupling, these experimental data could tell us
that the doped magnet is undergoing a transition of the same nature of that
between the renormalized classical and the QCR of the 2DQNL$\sigma$M. This
reasoning is still controversial and we will come back to this point at the
end of Sec.~\ref{s.expt}.

\section{The Effective Hamiltonian}
\label{s.eh}

The effective Hamiltonian approach has been successfully applied in the
last decade in the study of many different physical problems (for an
extensive review see Ref.~\onlinecite{CGTVV95}); the method is based on the
path-integral formalism and allows us to express the quantum statistical
average of physical observables in the form of classical-like phase-space
integrals. Other methods in quantum statistical mechanics formally lead to
this same form (see, e.g., Ref.~\onlinecite{Lee95}), but they differ from
each other because of the different approximations used to determine the
effective phase-space density.

In the effective Hamiltonian method developed for a flat phase space, such
approximation is the pure-quantum self-consistent harmonic
approximation, whose name, though long, is at least self-explanatory. The
PQSCHA does in fact separate the classical from the pure-quantum
contribution to the thermodynamics of the system, and then approximates
only the latter at a self-consistent harmonic level. This means that the
classical physics is exactly described at any temperature, and so are the
purely quantum linear effects, as the self-consistent harmonic
approxiamtion (SCHA) only affects the pure-quantum nonlinear contribution.
In other words, we do not renounce the exact description of the classical
behaviour in its full nonlinearity just because we cannot deal with the
nonlinear quantum corrections to it; we rather approximate only the latter.
This result is specially valuable in the study of magnetic systems because
their behaviour is very often characterized by long-wavelength excitations
(such as solitons, vortices or Goldstone modes), whose character is indeed
essentially classical.

The procedure leading to the effective Hamiltonian in the magnetic
case\cite{CTVV92epfc} can be briefly summarized as follows. First of all,
the spin Hamiltonian must be written in a bosonic form through a properly
chosen spin-boson transformation. Secondly, the Weyl symbol\cite{Berezin80}
of the resulting bosonic Hamiltonian has to be determined and the quantum
renormalizations\cite{CTVV92ham} made explicit. Finally, the resulting
effective Hamiltonian must be put into the form of a classical spin
Hamiltonian (by the inverse of the classical counterpart of the spin-boson
transformation used in the first step). The specific case of the 2DQHAF is
described in Appendix \ref{a.esh}, where the detailed derivation of the
effective spin Hamiltonian, that was just
sketched in Ref. \onlinecite{CTVV96prl}, is reported.

In choosing the spin-boson transformation for the isotropic
Hamiltonian Eq.~(\ref{e.ham}), we exclude the Villain transformation which
is designed for models with easy-plane anisotropy. Furthermore, neither
the Holstein-Primakoff\cite{HolsteinP40} (HP) nor the
Dyson-Maleev\cite{Dyson-Maleev} (DM) transformation
apparently gives an alternative, as they both break the symmetry of the
problem. However, the broken symmetry of the (assumed ordered) ground state
of the 2DQHAF is restored at finite temperature by long-wavelength
excitations whose effect, as mentioned above, is almost entirely taken into
account already at the classical level. As a consequence, it is almost
entirely taken into account also by the PQSCHA, no matter what spin-boson
transformation is used to evaluate the quantum renormalizations, as far as
the bosonic effective Hamiltonian is eventually put into the form of a
classical spin Hamiltonian. This means that by using the DM or HP
transformation in the PQSCHA framework, we break the $O(3)$ symmetry,
but only as far as the pure-quantum fluctuations are concerned, so that all
the essential features due to the symmetry of the Hamiltonian are actually
kept and the use of such transformations is consequently justified.
As for the choice between HP and DM, we employed the latter in Ref.
\onlinecite{CTVV96prl}; in Appendix~\ref{a.sbt} we show them to be
equivalent in this context.

An important point is the ordering problem. Whenever a theory
prescribes a function or functional to be associated with a quantum
operator, it should also specify the ordering rule to be used for that
purpose, to avoid an unnecessary uncertainty to enter the theory.
When dealing with spin systems, such uncertainty often manifests itself in
the ambiguous definition of the spin length, leading to an arbitrary
choice between $S$,$\sqrt{S(S+1)}$, or others.
The PQSCHA gives an unambiguous response to this point, by asking for the
Weyl symbol of the quantum operator to enter the formulae for its own
renormalization and, as we will see later, it makes the {\it effective spin
length}
\begin{equation}
 \widetilde{S}=S+{1\over2}
\end{equation}
naturally appear \cite{CTVV96prl} (see Appendix \ref{a.sbt}). Consistently
with its being the spin length in this formalism, $\widetilde{S}$ sets the
energy scale through the combination $J\widetilde{S}^2$. We therefore
define the {\it reduced temperature} as
\begin{equation}
 t \equiv {T\over J\widetilde{S}^2}~.
\end{equation}

The final result for the effective spin Hamiltonian is
\begin{eqnarray}
 {{\cal H}_{\rm{eff}}\over J\widetilde{S}^2}
 &=& -{\theta^4\over2} \sum_{{\bf{i}},{\bf{d}}} {\mbox{\boldmath$s$}}_{\bf{i}}
 {\cdot} {\mbox{\boldmath$s$}}_{{\bf{i}}+{\bf{d}}} + N\,{\cal G}(t)~,
\label{e.Heff}
\\
 {\cal{G}}(t) &=& {t\over N}\sum_{\bf{k}}\ln{\sinh{f_k}\over\theta^2f_k}
 - 2 \kappa^2 {\cal{D}}~,
\end{eqnarray}
with the temperature and spin dependent parameters
\begin{eqnarray}
 \theta^2&=&1-{{\cal D}\over2}~,
\label{e.theta}
\\
 {\cal D}&=&{1\over\widetilde S~ N} \sum_{\bf{k}}
 (1-\gamma_{\bf{k}}^2)^{1/2}
 \Big(\coth f_{\bf{k}}-{1\over f_{\bf{k}}}\Big)~,
\label{e.D}
\\
f_{\bf{k}}&=&{\omega_{\bf{k}}\over2\widetilde{S}t}~;
\label{e.fk}
\end{eqnarray}
moreover, $\gamma_{\bf{k}}=(\cos{k_1}+\cos{k_2})/2$, $N$ is the number of
sites of the lattice and ${\bf{k}}\equiv(k_1,k_2)$ is the wave vector in
the first Brillouin zone. ${\cal{D}}$ is the {\it pure-quantum}
renormalization coefficient, which takes the main contribution from the
high-frequency part (short-wavelength) of the spin-wave spectrum, because
of the appearance of the Langevin function.

As for the frequencies $\omega_{\bf{k}}$, note that in the PQSCHA context
they just appear in the evaluation of the pure-quantum renormalizations. In
the general PQSCHA they depend upon the phase-space coordinate, but such
dependence, when the system has many degrees of freedom, would make the
evaluation of the phase-space integrals with the effective Hamiltonian a
task as time demanding as a quantum MC simulation. Therefore, in deriving
the above formulas, we have actually introduced a low-coupling
approximation (LCA) to make the PQSCHA frequencies configuration
independent (see Appendix \ref{a.esh}). On the other hand, since the
effective Hamiltonian is affected by an approximate evaluation of the
renormalized frequencies just at a secondary level, the approximation is
worthy and assures the final results to depend only weakly on the specific
LCA used.

There are in fact several ways of defining a possible LCA in this context.
In Ref.~\onlinecite{CTVV92ham}, the fundamental phase-space-dependent
parameters appearing in the theory (in terms of which the frequency are
defined), say $A({\mbox{\boldmath$p$}},{\mbox{\boldmath$q$}})$, were
approximated in the simplest way as
$A({\mbox{\boldmath$p$}},{\mbox{\boldmath$q$}})\simeq{A}
\equiv{A}({\mbox{\boldmath$p$}}_0,{\mbox{\boldmath$q$}}_0)$; however, in the
specific case of the 2DQHAF the minimum energy configuration becomes
unstable as soon as the temperature is switched on, so that a more refined
LCA is in fact necessary for a proper description of the low-temperature
regime. In Ref.~\onlinecite{CTVV92ham} we suggested for this case that
$A({\mbox{\boldmath$p$}},{\mbox{\boldmath$q$}})$ could be better
approximated with its self-consistent average defined by the effective
Hamiltonian, i.e.,
\begin{equation}
 A({\mbox{\boldmath$p$}},{\mbox{\boldmath$q$}})\simeq
 {1\over{\cal Z}}
 \int d{\mbox{\boldmath$p$}} d{\mbox{\boldmath$q$}}
 ~A({\mbox{\boldmath$p$}},{\mbox{\boldmath$q$}})
 ~e^{-\beta{\cal H}_{\rm{eff}}}
 \equiv \langle
 ~A({\mbox{\boldmath$p$}},{\mbox{\boldmath$q$}})
 \rangle_{\rm{eff}}~,
\end{equation}
where ${\cal{Z}}$ is the partition function relative to
${\cal{H}}_{\rm{eff}}$. However, if an analytical expression for such
averages is not available, this approximation does not actually lighten the
burden of the numerical work. The best one can do analytically is to
evaluate them in the framework of the classical SCHA. In Appendix~\ref{a.esh}
we show that this leads to two coupled equations that give us all the
ingredients we need to actually use the effective Hamiltonian,
\begin{eqnarray}
 \omega_{\bf{k}}&=&4 \kappa^2 (1-\gamma^2_{\bf k})^{1/2}~;
\label{e.omegasc}
\\
 \kappa^2&=&1-{1\over2}\big({\cal D}+{\cal D}_{\rm{cl}}\big)
 =\theta^2-{{\cal D}_{\rm{cl}}\over2}~,
\label{e.kappasc}
\end{eqnarray}
where ${\cal{D}}_{\rm{cl}}=t/2\kappa^2$ represents the contribution to the
frequency renormalization due to the classical part of the fluctuations; at
variance with the pure-quantum coefficient ${\cal{D}}$, which is a
decreasing function of temperature, ${\cal{D}}_{\rm{cl}}$ rises with $t$.
As for the solution of the above equations, $\kappa^2(t)$ is real and
positive for $t{\le}\theta^4$, but becomes unphysically complex for
$t>\theta^4$. From the explicit solution
\begin{equation}
 \kappa^2={1\over2}\,\big[\theta^2+(\theta^4-t)^{{1\over2}}\big]
\end{equation}
one sees the instability to be originated just by the classical
contribution to $\kappa^2$.
This instability is typical of the SCHA and it is now
shared by the PQSCHA because of the particular type of LCA chosen
in order to optimize the low-temperature results. A closer look to
Eq.~(\ref{e.kappasc}) shows that what causes the instability is the
contribution to the frequency renormalization due to spin waves with long
wavelength. On the other hand, linear excitations with
$\lambda{\gtrsim}2\xi$ are in fact unphysical as they do not exist in a
system with no LRO; in their stead the system develops nonlinear
excitations which are fundamental in determining the thermodynamics of the
system, but whose contribution to the frequency renormalization is seen to
be negligible by the same arguments showing them to be responsible for the
vanishing of the magnetization.

Although the instability only affects the evaluation of the renormalized
frequencies, we want to devise a reasonable way to treat it, in order to
optimize the description of the temperature region where the above
mentioned nonlinear excitations become relevant in the thermodynamics of
the system, which is in fact the same region where the instability occurs.
The unphysical contribution in Eq.~(\ref{e.kappasc}) is then pulled up by
inserting a cutoff $|{\bf k}|\gtrsim\pi/\xi$, over the AFM Brillouin zone,
in the evaluation of ${\cal{D}}_{\rm{cl}}$. Such a cutoff is obviously
$t$ dependent, being $\xi=\xi(t)$, and it is relevant in the temperature
region where quantum nonlinear excitations are most important in the
system. As we will see in Sec.~\ref{s.expt}, such region seems to
coincide with that where other authors devised anomalous behaviours,
ascribed to the occurrence of quantum criticality in the theoretical
works\cite{ChubukovSY94}, or to doping effects in the experimental
ones\cite{CarrettaEtal97}.

Coming back to the effective Hamiltonian (\ref{e.Heff}), we see the quantum
effects to cause ({\it{i}}) the appearance of the spin length
$\widetilde{S}$; ({\it{ii}}) the renormalization of the exchange integral
$J$ by the factor $\theta^4(t)<1$; ({\it{iii}}) the introduction of a
uniform term ${\cal{G}}(t)$.

As far as ${\cal{G}}(t)$ is concerned, although it does not play any role
in calculating thermal averages, it contains the essential logarithmic term
that transforms the spin-wave contribution to the free energy from
classical to quantum\cite{CTVV92ham}. At $t=0$, ${\cal{G}}(0)=0$ and
the zero-temperature energy per site is $-2\theta^4(0)$, higher than the
classical value $-2$, because of the zero-point quantum fluctuations.
We have already commented on the appearance of $\widetilde{S}$ as a
consequence of a precise ordering prescription; we just add here that this
is a genuine quantum effect, being the noncommutativity of quantum
operators the only reason for an ordering rule to be necessary.

\begin{figure}[hbt]
\centerline{\psfig{bbllx=16mm,bblly=69mm,bburx=192mm,bbury=207mm,%
figure=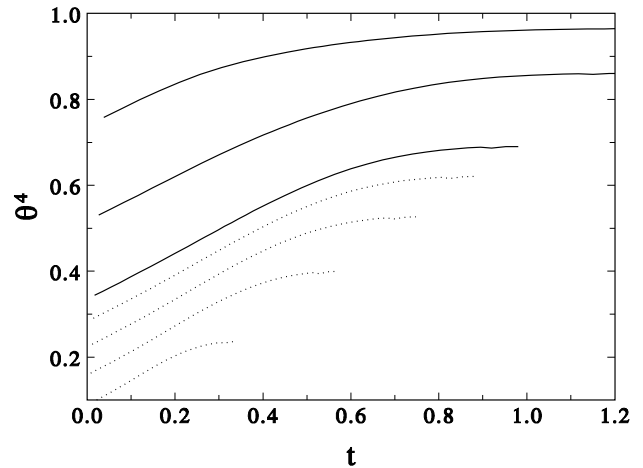,width=82mm,angle=0}}
\caption{
Renormalization parameter $\theta^4$ vs $t$, for (from the top curve)
$S=5/2$, $1$, $1/2$ (solid lines), and $S=0.4$, $0.3$, $0.2$, $0.1$
(dotted lines).
\label{f.theta}
}
\end{figure}

Let us now examine the exchange-integral renormalization embodied in the
factor $\theta^4$; the parameter $\theta^2$ depends on both $S$ and $t$, it
is well defined as far as a physical solution for $\kappa^2$ is available;
it attains its minimum value for $t=0$,
where $\theta^2=\kappa^2$ coincides with the one-loop correction to the
spin-wave velocity. For increasing $S$ or $t$, $\theta^2$ increases, going
asymptotically to $1$ in the classical ($S{\rightarrow}\infty$) or
high-temperature ($t{\rightarrow}\infty$) limit. The instability value
$\theta^2=0$ is not reached for physical values of the spin, being
$\theta^2>1{-}(2\widetilde{S})^{-1}>0$ for $S>0$.

\begin{figure}[hbt]
\centerline{\psfig{bbllx=16mm,bblly=69mm,bburx=192mm,bbury=207mm,%
figure=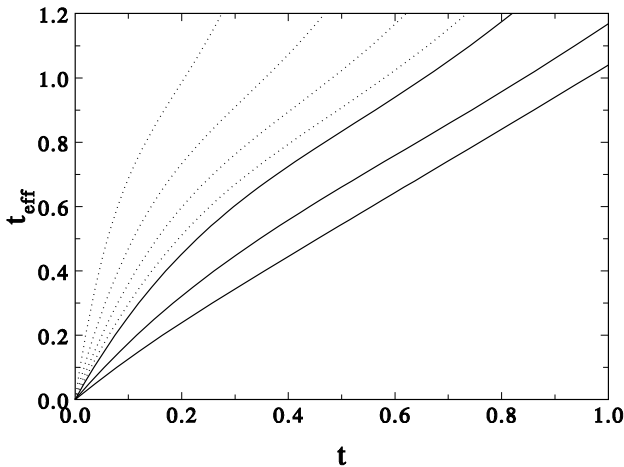,width=82mm,angle=0}}
\caption{
Effective classical temperature $t_{\rm{eff}}=t/\theta^4(t)$ vs $t$, for
(from the bottom curve) $S=5/2$, $1$, $1/2$ (solid lines) and $S=0.4$,
$0.3$, $0.2$, $0.1$ (dotted lines).
\label{f.teff}
}
\end{figure}

The essential information we get from Eq.~(\ref{e.Heff}) is that the 2DQHAF
at an actual temperature $t$ behaves as its classical counterpart at an
effective temperature
\begin{equation}
 t_{\rm{eff}}={t\over \theta^4(t)}~,
\label{e.teff}
\end{equation}
or, in other terms, that the energy scale is renormalized by a temperature
dependent factor $\theta^4(t)$. In Figs.~\ref{f.theta} and \ref{f.teff} we
show $\theta^4$ and $t_{\rm{eff}}$ as functions of $t$, for different
spins, including some $S<1/2$ unphysical values (dotted lines). As $S$
decreases the difference between $t$ and $t_{\rm{eff}}$ becomes more and
more pronounced, because of $\theta^2$ getting smaller, indicating larger
quantum fluctuations in the system.

Our theory is quantitatively meaningful just as far as the renormalization
coefficient ${\cal{D}}$ is small enough to justify the self-consistent
harmonic treatment of the pure-quantum effects; although this is not
obviously the case for the $S<1/2$ low-temperature regime, the dotted
lines at least qualitatively suggest that no critical behaviour occurs, no
matter how small the spin value. What we rather see is that the sharp
dependence of $t_{\rm{eff}}$ upon $t$ brings to lower temperatures
those features which are indeed typical of the highly disordered,
high-temperature regime of the classical model.
In Sec.~\ref{s.expt.dop} we will come back to this point, in relation
with the effects of magnetic doping, and we just recall here that any
reasoning about models with $S<1/2$ should just be considered as
speculative when dealing with real magnets.

\section{Thermodynamic properties}
\label{s.tp}

Once the effective Hamiltonian has been determined, the thermodynamic
properties can be derived from the partition function
${\cal{Z}}=\int{d}{\mbox{\boldmath$p$}}
{d}{\mbox{\boldmath$q$}}\exp(-\beta{\cal{H}}_{\rm{eff}})$; more general
statistical averages are given by
\begin{equation}
 \big\langle\hat {\cal O}\big\rangle={1\over {\cal Z}}
 \int d^{\scriptscriptstyle N}\!{\mbox{\boldmath$s$}}
 ~\widetilde{\cal O}~
 e^{-\beta {\cal H}_{\rm{eff}}}
 \equiv\big\langle\widetilde{\cal{O}}\big\rangle_{\rm{eff}}
\label{e.aveO}
\end{equation}
where
$\widetilde{\cal{O}}\equiv\widetilde{\cal{O}}
(\{{\mbox{\boldmath$s$}}_{\bf{i}}\})$
is obtained by the quantum operator $\hat O$
following the same procedure used to determine the configurational part of
${\cal{H}}_{\rm{eff}}$ (see Appendix~\ref{a.esh}); by
$\langle\cdots\rangle_{\rm{eff}}$ we hereafter mean the classical thermal
average with the effective Hamiltonian, which equals the classical average
at the effective temperature $t_{\rm{eff}}$.

In order to evaluate the classical 2D phase-space integrals appearing in
$\langle\cdots\rangle_{\rm{eff}}$ according to Eq.~(\ref{e.aveO}), a
standard classical MC simulation is perfectly suitable; being the
existent classical MC data\cite{ShenkerT80,Kim94} incomplete as far as the
correlation functions are concerned, we have performed our own simulations
(following the same procedure described in Ref.~\onlinecite{CTV95}, with
slight modifications of the overrelaxed moves made to account for the fully
isotropic exchange) on a $256\times{256}$ square lattice, for
$0.54{\le}t{\le}1.42$. The dimension of the lattice ensures the data to be
free from saturation effects, according to the $L{\gtrsim}6\xi$ criterion,
for all temperatures but the first two (0.54 and 0.56).

\begin{figure}[hbt]
\centerline{\psfig{bbllx=16mm,bblly=69mm,bburx=192mm,bbury=207mm,%
figure=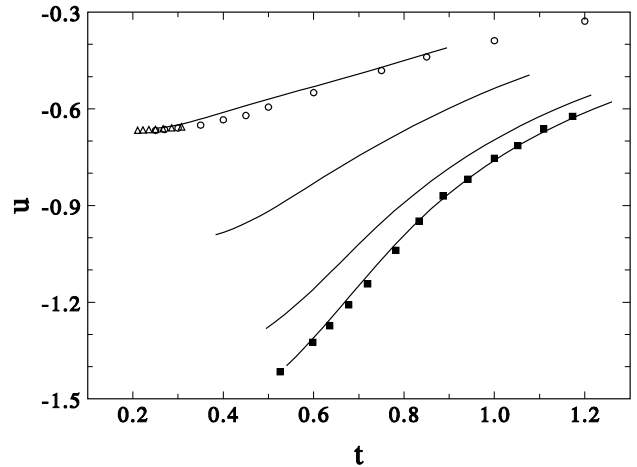,width=82mm,angle=0}}
\caption{
Energy per spin $u=\langle\hat{\cal{H}}\rangle/(NJ\widetilde{S}^2)$ vs $t$,
for (from the bottom curve) $S=\infty$, $5/2$, $1$ and $1/2$; the symbols
are quantum MC data for $S=1/2$ (circles\protect\cite{MakivicD91} and
triangles\protect\cite{KimLT97}) and previous classical MC
data\protect\cite{ShenkerT80} (squares).
\label{f.inten}
}
\end{figure}

\subsection{Internal energy and specific heat}
\label{s.tp.iesh}

According to Eq.~(\ref{e.aveO}), the internal energy per spin
$u\equiv\langle\hat{\cal{H}}\rangle/(NJ\widetilde{S}^2)$ is
\begin{equation}
 u(t)=\theta^4(t)~u_{\rm{cl}}(t_{\rm{eff}})~,
\label{e.inten}
\end{equation}
where $u_{\rm{cl}}$ is the classical energy;
from the results shown in Fig.~\ref{f.inten} we see the quantum
renormalizations to increase the energy and flatten the curve $u(t)$ at all
temperatures and more markedly for smaller spin. Quantum fluctuations are
responsible for both effects, as they introduce additional and almost
temperature-independent disorder, thus making the system more unstable,
i.e., with many different configurations thermodynamically relevant, already
at low temperatures.

\begin{figure}[hbt]
\centerline{\psfig{bbllx=16mm,bblly=69mm,bburx=192mm,bbury=207mm,%
figure=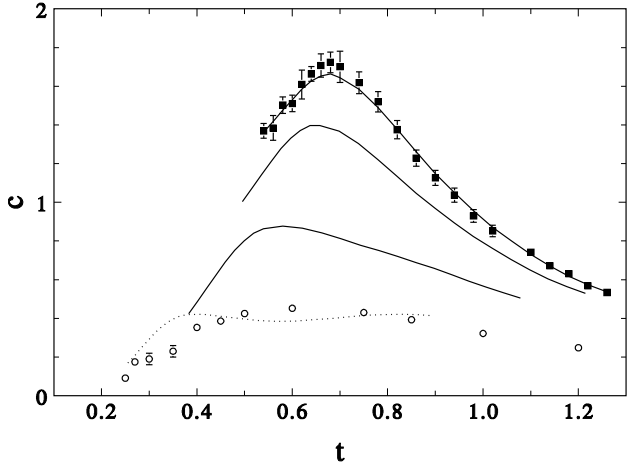,width=82mm,angle=0}}
\caption{
Specific heat per spin $c\equiv\partial{u}/\partial{t}$ vs $t$,
for (from the top curve) $S=\infty$, $5/2$, $1$ and $1/2$;
circles as in Fig.~\protect\ref{f.inten}; squares are our MC results.
For the (dotted) $S=1/2$ curve see comments in the text.
\label{f.sh}
}
\end{figure}

Consistently with this picture, Fig.~\ref{f.sh} shows the peak of the
specific heat moving towards lower temperatures and decreasing in height,
as $S$ decreases. Our data for the specific heat at finite spin are
obtained by numerical derivation of $u(t)$, as given by Eq.~(\ref{e.inten});
because of the temperature dependence of the cutoff involved in the LCA
(as described in Sec.~\ref{s.eh}), they are affected by consistent
numerical uncertainty and should not be considered but qualitatively for
$S=1/2$ where such dependence is more pronounced.

\subsection{Correlation functions}
\label{s.tp.cf}

For the correlation functions
$G({\bf{r}})\equiv\langle\hat{\mbox{\boldmath$S$}}_{\bf{i}}
{\cdot}\hat{\mbox{\boldmath$S$}}_{{\bf{i}}+{\bf{r}}}\rangle$, with
${\bf{r}}\equiv(r_1,r_2)$ any vector on the square lattice, we find (see
Appendix~\ref{a.esh})
\begin{equation}
 G({\bf{r}})= \widetilde{S}^2 \theta^4_{\bf{r}}
 \langle{\mbox{\boldmath$s$}}_{\bf{i}}
 {\cdot}{\mbox{\boldmath$s$}}_{{\bf{i}}+{\bf{r}}}\rangle_{\rm{eff}}~,
\label{e.Gdr}
\end{equation}
where $\theta^4_{{\bf{r}}}=1{-}{1\over2}{\cal{D}}_{\bf{r}}$ and the
renormalization parameter ${\cal{D}}_{\bf{r}}$ reads
\begin{equation}
 {\cal D}_{\bf{r}}={1\over\widetilde SN} \sum_{\bf{k}}
 \bigg( { 1+\gamma_{\bf{k}} \over 1-\gamma_{\bf{k}} } \bigg)^{1/2}
 \Big(\coth f_{\bf{k}}{-}{1\over f_{\bf{k}} } \Big)
 (1{-}\cos{\bf{k}}{\cdot}{\bf{r}})~;
\label{e.Dr}
\end{equation}
note that for nearest neighbors, ${\bf{r}}={\bf{d}}$, we have
${\cal{D}}_{\bf{d}}\equiv{\cal{D}}$.

\begin{figure}[hbt]
\centerline{\psfig{bbllx=16mm,bblly=69mm,bburx=192mm,bbury=207mm,%
figure=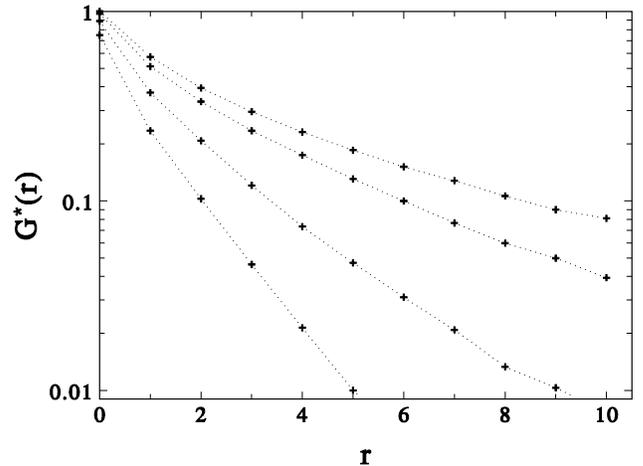,width=82mm,angle=0}}
\caption{
Absolute value of the spin correlation function
$G^*(r)=|G(r,0)|/\widetilde{S}^2$ vs $r$, at $t=0.7$ and for (from
the top curve) $S=\infty$, $5/2$, $1$, and $1/2$. The dotted lines are just
guides for the eye.
\label{f.corall}
}
\end{figure}

Equation (\ref{e.Gdr}) shows that the quantum correlation functions can be
obtained by multiplying the classical ones at the effective temperature
$t_{\rm{eff}}$ by the temperature and spin-dependent renormalization factor
$\theta^4_{{\bf{r}}}\widetilde{S}^2$. The overall effect is that of a
strong reduction of the spin correlations, as it is clear from
Fig.~\ref{f.corall}, where
$G^*(r)\equiv\big|G(|{\bf{r}}|)\big|/\widetilde{S}^2$ is shown for
${\bf{r}}=(r,0)$, at a fixed temperature and for different values of the
spin.

\begin{figure}[hbt]
\centerline{\psfig{bbllx=16mm,bblly=69mm,bburx=192mm,bbury=207mm,%
figure=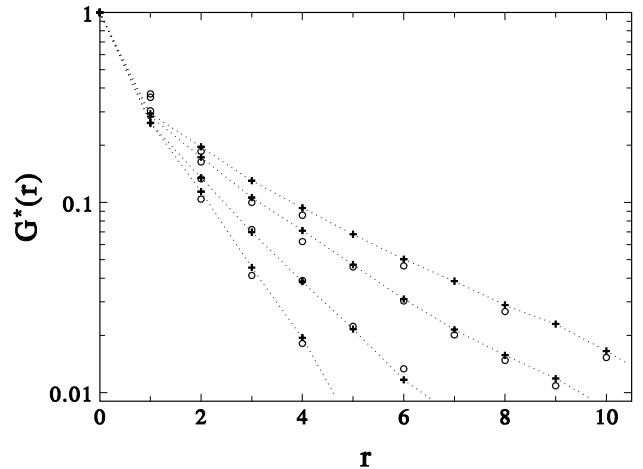,width=82mm,angle=0}}
\caption{
Absolute value of the spin correlation function
$G^*(r)=|G(r,0)|/\widetilde{S}^2$ vs $r$, at $S=1/2$ and for (from
the top curve) $t=0.45$, $0.50$, $0.60$, $0.75$; also reported are quantum
MC data \protect\cite{MakivicD91} (circles). The dotted lines are just
guides for the eye.
\label{f.corS05}
}
\end{figure}

In order to exploit some available quantum MC data for $S=1/2$, we
have performed four classical MC simulations with the effective
Hamiltonian, at the temperatures corresponding, via Eq.~(\ref{e.teff}), to
those at which quantum MC simulations have been performed by Makivi\`c and
Ding\cite{MakivicD91}. As seen in Fig.~\ref{f.corS05}, the agreement we
find is very good, and gives us confidence to proceed towards the
evaluation of the susceptibility and of the correlation length, both indeed
deriving from the correlation functions.

\section{Facing the experimental data}
\label{s.expt}

In this last section we compare our results with the available experimental
data and show that, at variance with what seems to emerge from the analysis
based on the QNL$\sigma$M approach, there is no substantial difference
between the agreement we find for systems with $S=1/2$, and those with
$S\geq{1}$.

We consider two fundamental physical observables, the staggered
susceptibility $\chi$ and the correlation length $\xi$, and we underline
once more that our results do not contain free parameters and need nothing
but the values of $J$ and $S$ to be compared with the experimental data. In
the case of the staggered susceptibility, the latter come in arbitrary
units so that a multiplicative factor must be determined by the standard
least-squares procedure.
We also remind that our results for $S{=}1/2$ and low temperatures
($t{\lesssim}0.35$) are obtained in a region where the PQSCHA, given the
relatively large value of the renormalization coefficient ${\cal{D}}$,
touches its limit of validity\cite{CTVV97prlre}.

\begin{figure}[hbt]
\centerline{\psfig{bbllx=16mm,bblly=69mm,bburx=192mm,bbury=207mm,%
figure=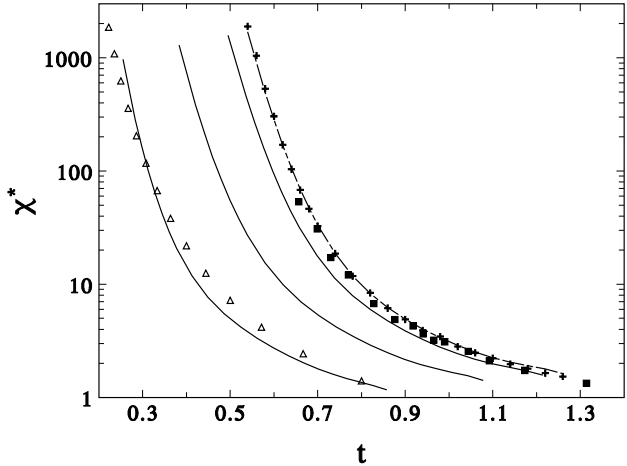,width=82mm,angle=0}}
\caption{
Staggered susceptibility $\chi^*=\chi/\widetilde{S}^2$ vs $t$, for (from
the rightmost curve) $S=\infty$, $5/2$, $1$, and $1/2$; the crosses (our
simulations, see text) and the squares\protect\cite{ShenkerT80} are
classical MC data, the triangles are quantum MC
data\protect\cite{KimLT97} for $S=1/2$.
\label{f.chiall}
}
\end{figure}

\subsection{The staggered susceptibility}
\label{s.expt.chi}

The staggered susceptibility for the 2DQHAF is defined as
\begin{equation}
 \chi={1\over3}\sum_{{\bf{r}}}(-)^{r_1+r_2}~G({\bf{r}})
\end{equation}
and by the PQSCHA result Eq.~(\ref{e.Gdr}) we find
\begin{equation}
 \chi={1\over 3}\bigg[
 S(S+1) + \widetilde{S}^2 \sum_{{\bf{r}}\neq 0} (-)^{r_1+r_2}
 ~\theta^4_{{\bf{r}}} ~\langle{\mbox{\boldmath$s$}}_{\bf{i}}
 {\cdot}{\mbox{\boldmath$s$}}_{{\bf{i}}+{\bf{r}}}\rangle_{\rm{eff}} \bigg]~.
\label{e.chi}
\end{equation}
By using our classical MC data for $\langle{\mbox{\boldmath$s$}}_{\bf{i}}
{\cdot}{\mbox{\boldmath$s$}}_{{\bf{i}}+{\bf{r}}}\rangle_{\rm{eff}}$ we then
obtain the quantum results for all values of $S$, as the spin- and
temperature-dependent renormalization factor $\theta^4_{{\bf{r}}}$ can be
easily evaluated for any value of ${\bf{r}}$.

\begin{figure}[hbt]
\centerline{\psfig{bbllx=16mm,bblly=69mm,bburx=192mm,bbury=207mm,%
figure=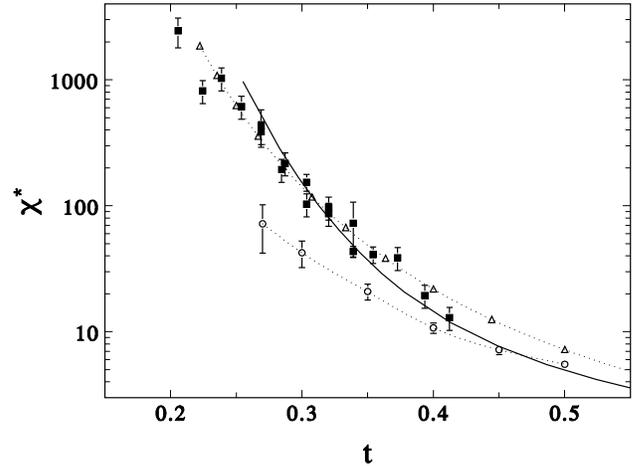,width=82mm,angle=0}}
\caption{
Staggered susceptibility $\chi^*=\chi/\widetilde{S}^2$ vs $t$, for $S=1/2$.
Circles\protect\cite{MakivicD91} and triangles\protect\cite{KimLT97}
are quantum MC data (the dotted lines are just guides for the eye); the
squares are neutron scattering data\protect\cite{GrevenEtal94} for
Sr$_2$CuO$_2$Cl$_2$.
\label{f.chiS05}
}
\end{figure}

\begin{figure}[hbt]
\centerline{\psfig{bbllx=16mm,bblly=69mm,bburx=192mm,bbury=207mm,%
figure=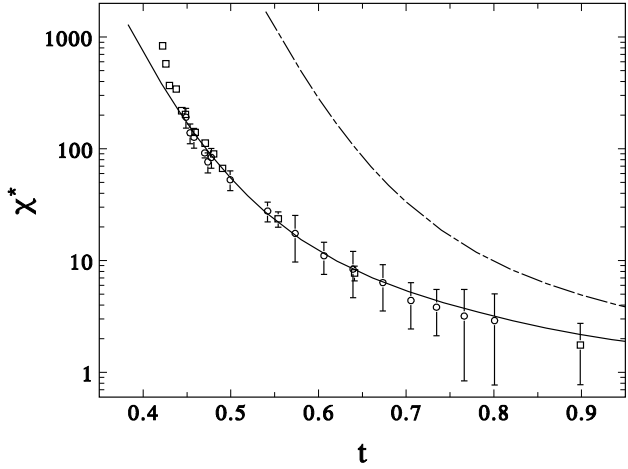,width=82mm,angle=0}}
\caption{
Staggered susceptibility $\chi^*=\chi/\widetilde{S}^2$ vs $t$, for $S=1$.
Circles and squares are neutron scattering
data for La$_2$NiO$_4$ (Ref.~\protect\onlinecite{NakajimaEtal95})
and K$_2$NiF$_4$ (Ref.~\protect\onlinecite{GrevenEtal94}), respectively.
The classical result (dash-dotted line) is also reported.
\label{f.chiS10}
}
\end{figure}

In Fig.~\ref{f.chiall} we show $\chi^*\equiv\chi/\widetilde{S}^2$ as a
function of temperature for different spin values. For $S=\infty$ we have
reported our MC data, as obtained from both the general definition
(\ref{e.chi}) (curve) and the classical expression
$\chi^*=\big\langle|\sum_{\bf{i}}{\mbox{\boldmath$s$}}_{\bf{i}}|^2
\big\rangle/3N$ (crosses); also reported are classical MC data from
Ref.~\onlinecite{ShenkerT80}, as well as quantum MC data recently obtained
by Kim {\it et~al.}\cite{KimLT97} for $S=1/2$. Figure~\ref{f.chiS05}
shows our results for $S=1/2$ together with  experimental
data\cite{GrevenEtal94} for the real compound Sr$_2$CuO$_2$Cl$_2$; the
quantum MC results by Makivi\`c and Ding are also shown.
The case $S=1$ is considered in Fig.~\ref{f.chiS10}, where the
experimental data\cite{NakajimaEtal95,GrevenEtal94}
for the compounds La$_2$NiO$_4$ and K$_2$NiF$_4$ are reported.
The agreement between our theoretical curves and the experimental data is
substantially equivalent in both the $S=1/2$ and $S=1$ case, strongly
suggesting that the difficulties encountered by a QNL$\sigma$M-based
analysis to draw a unique picture for different spin values, derive from
the inadequacy of the theory, rather than from an actual difference in the
thermodynamics of the different compounds.

\begin{figure}[hbt]
\centerline{\psfig{bbllx=16mm,bblly=69mm,bburx=192mm,bbury=207mm,%
figure=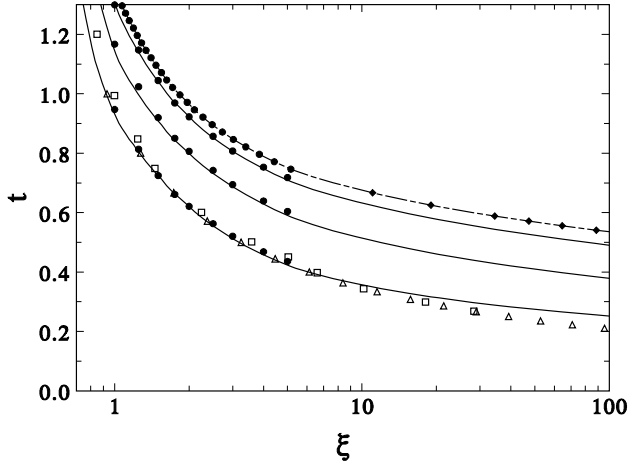,width=82mm,angle=0}}
\caption{
Curves $t(\xi)$ as explained in text, for (from the top curve)
$S=\infty$, $S=5/2$, $1$, and $1/2$,
the triangles\protect\cite{KimLT97} and the
squares\protect\cite{MakivicD91} are quantum MC data for $S=1/2$, the
circles are HTE results\protect\cite{Elstner97Etal95}, and
the diamonds are classical MC data\protect\cite{Kim94}.
\label{f.txi}
}
\end{figure}

\subsection{Correlation length}
\label{s.expt.xi}

The correlation length is defined by the asymptotic behaviour for
$r\equiv|{\bf{r}}|\to\infty$ of the correlation function,
$|G({\bf{r}})|\sim e^{-r/\xi}$~.
From the PQSCHA result, Eq.~(\ref{e.Gdr}), since for large $r$
$\theta^4_{\bf{r}}\to{\rm{const}}$, we get
$\langle{\mbox{\boldmath$s$}}_{\bf{i}}{\cdot}
{\mbox{\boldmath$s$}}_{{\bf{i}}+{\bf{r}}}\rangle_{\rm{eff}}\sim{e}^{-r/\xi}$~,
which means, comparing the definition (\ref{e.aveO}) and using
Eq.~(\ref{e.teff}),
\begin{equation}
 \xi(t)=\xi_{\rm{cl}}(t_{\rm{eff}})~,
\label{e.scalexi}
\end{equation}
i.e., the correlation length $\xi$ of the quantum model at a certain
temperature $t$ equals the classical one $\xi_{\rm{cl}}$
at the effective temperature $t_{\rm{eff}}>t$.
It is easy to see that the relation empirically extracted by Elstner {\it
et~al.}\cite{Elstner97Etal95} from their HTE results
(i.e., $\xi(S,T){\approx}\xi_{\rm{cl}}\big[T/JS(S{+}1)\big]$ for $S>1$ and
$T{\geq}JS$) is nothing but the high-$T$ and high-$S$ limit of
Eq.~(\ref{e.scalexi}).
Furthermore, Eq.~(\ref{e.scalexi}) represents the correct expression of the
link between the classical and the quantum discrete magnetic model, to be
compared with the one devised by CHN for the QNL$\sigma$M in the
renormalized classical regime, leading from the classical result by
Br\'ezin and Zinn-Justin\cite{BrezinZ76} to Eq.~(\ref{e.xiCHN}).

Equation~(\ref{e.scalexi}) allows us to obtain the quantum correlation length
from the classical one by a simple temperature scaling. In other terms, we
see that each value $\xi$ corresponds to a temperature $t(\xi,S)$ which is
different for different spin values. This point of view is taken in
Fig.~\ref{f.txi}, where we report our results for $t(\xi,S)$, together with
MC and HTE data. In the $S=1/2$ case, we find that
$t(\xi,1/2)/t(\xi,\infty)=\theta^4\gtrsim{0.5}$ for $\xi{\gtrsim}10$; as
already pointed out at the end of Sec.~\ref{s.eh}, such a small value of
$\theta^4$ makes the PQSCHA no more quantitatively reliable. Nevertheless,
the figure shows that the accuracy of our results for $t(\xi,S)$ at
$S=1/2$ and $\xi\simeq{100}$, as from the comparison with the quantum MC
data, is still better than $20\%$, which makes them qualitatively
meaningful even in the extreme quantum region\cite{CTVV97prlre}.
However, the inaccuracy in the ``renormalized'' temperature gives larger
inaccuracy in the inverse function, $\xi(t)$, due to its exponential
behaviour.

\begin{figure}[hbt]
\centerline{\psfig{bbllx=16mm,bblly=69mm,bburx=192mm,bbury=207mm,%
figure=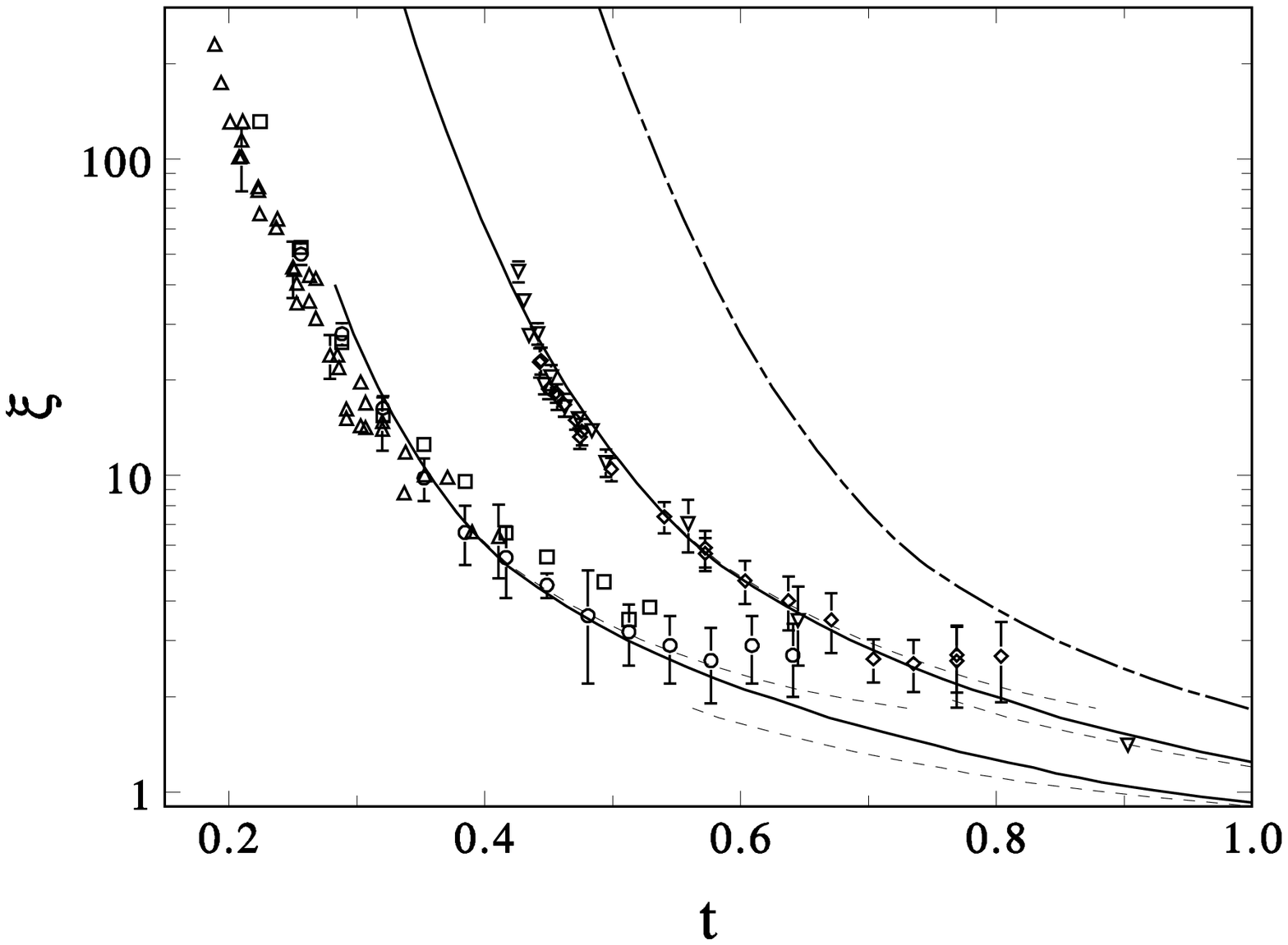,width=82mm,angle=0}}
\caption{
Correlation length $\xi$ vs $t$, for $S=1/2$ (leftmost) and $S=1$. The
symbols are experimental data; for $S=1/2$: $^{63}$Cu~NQR
data\protect\cite{CarrettaRS97} (circles) and neutron scattering data for
La$_2$CuO$_4$ (squares\protect\cite{BirgeneauEtalunpub97}) and for
Sr$_2$CuO$_2$Cl$_2$ (up-triangles\protect\cite{GrevenEtal94}); for $S=1$:
neutron scattering data for La$_2$NiO$_4$
(down-triangles\protect\cite{NakajimaEtal95}) and for K$_2$NiF$_4$
(diamonds\protect\cite{GrevenEtal94}). The classical result (dash-dotted
line) is also reported; the dashed lines are the low-$t$ and high-$t$
results of the PQSCHA (see text).
\label{f.xiS05-1}
}
\end{figure}

The experimental data for $S=1/2$ and $S=1$ are compared with our theory in
Fig.~\ref{f.xiS05-1}; at variance with what happens by using the
QNL$\sigma$M approach, the agreement does not get worse in the $S=1$ case.
For each spin we have also reported the curves (dashed lines) obtained by
using the standard LCA (high-$t$ curves) as well as the more accurate one
described in Sec.~\ref{s.eh} (low-$t$ curves) without cutoff and hence
truncated at $t=\theta^4$. They are smoothly connected by the continuous
curves, i.e., those obtained by using the cutoff. For each spin value, the
temperature interval where the cutoff is relevant identifies the region
where nonlinear excitations are present in the system, and hence suggests
where a possible crossover towards the QCR could occur.

\begin{figure}[hbt]
\centerline{\psfig{bbllx=16mm,bblly=69mm,bburx=192mm,bbury=207mm,%
figure=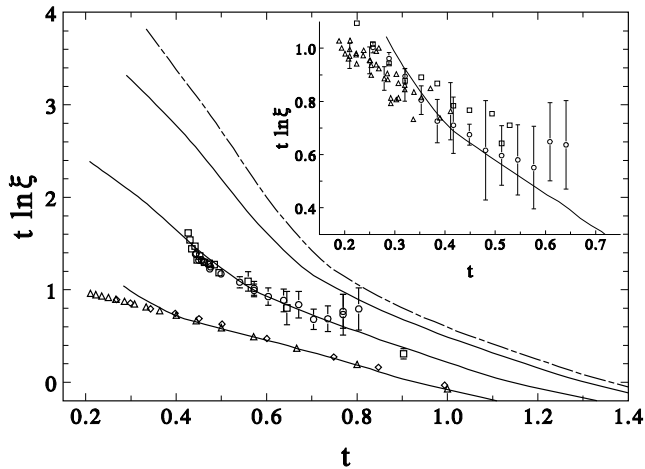,width=82mm,angle=0}}
\caption{
The function $y(t)=t\ln\xi$ vs $t$, for (from the rightmost curve)
$S=\infty$, $5/2$, $1$ and $1/2$; the triangles\protect\cite{KimLT97}
and the diamonds\protect\cite{MakivicD91} are quantum MC data for $S=1/2$;
also reported are neutron scattering data for La$_2$NiO$_4$
(circles\protect\cite{NakajimaEtal95}) and for K$_2$NiF$_4$
(squares\protect\cite{GrevenEtal94}). The inset reports, together with our
result for $S=1/2$ (line), $^{63}$Cu~NQR data
(circles\protect\cite{CarrettaRS97}) and neutron scattering data for
La$_2$CuO$_4$ (squares\protect\cite{BirgeneauEtalunpub97}) and for
Sr$_2$CuO$_2$Cl$_2$ (triangles\protect\cite{GrevenEtal94}).
\label{f.tlnxi}
}
\end{figure}

In order to better understand why the QNL$\sigma$M theory fails in
describing the 2DQHAF when used to interpret the available reference data
for $S\ge{1}$, the function $y(t)=t\ln\xi$ is shown in Fig.~\ref{f.tlnxi}
as a function of $t$. We remind the reader that the QNL$\sigma$M
approach gives $y_{\rm{q}}(t)=mt{+}n$ in the quantum case (with $m$ and
$n$ constant in temperature but $S$ dependent), being such expression based
on the classical two-loop result $y_{\rm{cl}}(t)=t\ln(\mu{t}){+}\nu$
(with $\mu$ and $\nu$ constants); first of all, one must hence check
whether or not this classical result can be safely extended to the magnetic
system.

Let us concentrate on the $S=\infty$ (classical) MC data reported in
Fig.~\ref{f.tlnxi} (rightmost curve): The complete curve cannot be fitted
with a function of the form $y_{\rm{cl}}$ and such fit is solely possible
if one restricts himself to a limited temperature range, as in the
intermediate-temperature region there is a clear change in the slope. On
the other hand, this should not surprise, as $t\ln\xi=t\ln(\mu t)+\nu$ is
a two-loop result and it is hence bound to be correct only at lowest
temperatures. We now expect this problem to propagate to the quantum case,
which is based on the classical result.

Let us then look at the $S=5/2$ and $S=1$ curves: according to the
QNL$\sigma$M these should be straight lines, but they are not; actually one
can easily see a change in the slope at intermediate temperature, followed
by a curvature inversion at lower $t$. As in the classical case, hence,
good fits with the function $y_{\rm{q}}$ can be obtained either in the low-
or (misleadingly) in the high-temperature region, but not on the whole
temperature range. To fit the experimental data for $S=1$ with a straight
line is seemingly impossible, as pointed out by several
authors\cite{Elstner97Etal95,GrevenEtal94,NakajimaEtal95} in the past few
years.

Finally, if we look at the $S=1/2$ case it becomes clear why the
QNL$\sigma$M approach gave such a good agreement when first used to fit the
experimental data. The change in both the slope and the curvature of
$t\ln\xi$ is less pronounced and possibly occurs at lower temperatures, the
lower the spin: in the $S=1/2$ case, we find difficult to say whether these
features are still present or not, but, if yes, they occur in a temperature
region where the extremely high value of $\xi$ ($\approx 10^4$) makes both
the experimental and the simulation data more difficult to be obtained. The
experimental data, as well as our results, do actually suggest a change in
the slope; on the other hand quantum MC data by Kim {\it
et~al.}\cite{KimLT97}, do not give evidence of such change.

As for the very-low temperature data by Suh {\it et~al.}\cite{SuhEtal95},
it is to be noticed that the authors explained the change in the slope of
their curve in terms of a crossover from an isotropic towards an easy-plane
(Kosterlitz-Thouless-like) behaviour. We cannot question their
interpretation, but we underline that their conclusion is based on the
assumption that the $t\ln\xi$ curve for the 2DQHAF is a straight line,
which is, as we are seeing, at least questionable.

\subsection{Magnetic doping and quantum criticality}
\label{s.expt.dop}

We have already pointed out that at intermediate temperatures there is an
interval, whose width is larger the smaller the spin, where quantum
nonlinear effects (due to higher order terms in the coupling) are
significant; such interval is identified, in our theory, by the temperature
region where the cutoff is relevant, a region which is very easily
recognizable in Fig.~\ref{f.xiS05-1} ($0.45\lesssim{t}\lesssim{0.9}$ for
$S=1/2$ and $0.65\lesssim{t}\lesssim{1}$ for $S=1$) and seems indeed to
coincide with that where Chubukov and Sachdev\cite{ChubukovS94} suggested
the occurrence of the QCR in the QNL$\sigma$M.

\begin{figure}[hbt]
\centerline{\psfig{bbllx=16mm,bblly=69mm,bburx=192mm,bbury=207mm,%
figure=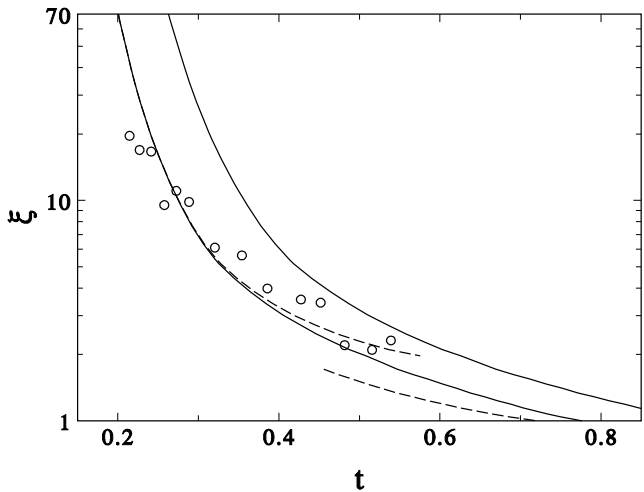,width=86mm,angle=0}}
\caption{
Experimental data for the correlation length,
for the magnetically doped compound La$_2$Cu$_{1{-}x}$Zn$_x$O$_4$,
from Ref.~\protect\onlinecite{CarrettaEtal97}.
The leftmost lines refer to $S=0.35$, with the dashed lines as in
Fig.~\protect\ref{f.xiS05-1}; for comparison the $S=1/2$ curve is also
reported (rightmost curve).
\label{f.xidop}
}
\end{figure}

Experimental data for magnetically doped materials have been interpreted by
several authors in terms of a possible crossover from the classical
renormalized towards the QCR. Let us concentrate, in particular, on the
data by Carretta {\it et~al.}\cite{CarrettaEtal97}, obtained by a scaling
analysis of their $^{63}$Cu nuclear quadrupole relaxation (NQR) data, for
the correlation length of La$_2$CuO$_4$ doped with nonmagnetic impurities,
i.e., for the compound La$_2$Cu$_{1-x}$Zn$_x$O$_4$. In Fig.~\ref{f.xidop} we
report their data for $x=0.018$: it is evident that the doping causes a
strong reduction of the spin correlation, as well as a clear flattening of
$\xi$ as a function of temperature. Such flattening does occur in the same
intermediate temperature region where the QCR has been suggested to occur.

The authors have interpreted their data in terms of an effective reduction
of the spin stiffness, consistently with our theory which shows similar
effects to be caused by an effective reduction of the spin value. The
curve in Fig.~\ref{f.xidop} is obtained by the PQSCHA with $S=0.35$, a
value which has been empirically determined in order to optimize the
agreement with the experimental data. Although the use of our approach for
$S<1/2$ is not fully justified, these results do at least qualitatively
suggest that for $S<1/2$ the magnetic model moves towards a regime where
nonlinear effects become relevant in a wider temperature region (the one
where $\xi$ shows a clear plateau) and such behaviour could be seen as a
signature of the crossover towards a QCR.

\section{Conclusions}
\label{s.concl}

In this paper we have applied the pure-quantum self-consistent
harmonic approximation \cite{CTVV92ham} (PQSCHA) to the study of the
thermodynamics of the two-dimensional Heisenberg antiferromagnet on the
square lattice. The PQSCHA allowed us to reduce the evaluation of quantum
averages to the calculation of classical-like phase-space integrals.
Therefore, using classical MC simulations, we have been able to obtain
results for several quantum thermodynamic quantities. What is remarkable is
that these results are fully determined by the system's parameters, i.e.,
the spin value $S$ and the ratio between the temperature and the exchange
energy constant $T/J$.

The main effect is seen to be the temperature-depen\-dent weakening of the
effective classical exchange constant, so that an effective classical
temperature naturally arises, Eq.~(\ref{e.teff}); this leads to a very
simple relation, Eq.~(\ref{e.scalexi}), directly giving the quantum
correlation length in terms of its classical counterpart. For other
thermodynamic quantities, such as correlation functions and the staggered
susceptibility, the expressions involve further pure-quantum
renormalization factors that can be straightforwardly computed.
Even pushing down the theory to the extreme quantum case, $S=1/2$,
we find agreement both with quantum MC and experimental data, until the
renormalization parameters of the theory become too large
for $T/J\lesssim{0.35}$ making the results of only qualitative value.

For higher spin, $S\geq{1}$, the theory is reliable at any temperature,
and it agrees indeed with the available experimental data, while quantum MC
simulations have not yet been feasible. It would be interesting to compare
also with the announced\cite{Elstner97Etal95} experimental results for the
$S=5/2$ square-lattice antiferromagnet Rb$_2$MnF$_4$, which are not yet
available.

\acknowledgments

We acknowledge hospitality at the ISIS Facility of the Rutherford Appleton
Laboratory (U.K.), where part of this work and most of the related classical
MC simulations were performed, and we thank S.~W. Lovesey, head of the
ISIS theory division.
V.~T. acknowledges fruitful exchanges of visits and discussions with
F. Borsa and A.~V. Chubukov.
We thank A. Rigamonti and P. Carretta, for useful discussion and for
providing us with experimental data, and J.-K. Kim, D.~P. Landau, and M.
Troyer, for their useful comments and for supplying unpublished quantum MC
data.

\appendix

\section{Spin-boson transformation and Weyl ordering}
\label{a.sbt}

In this appendix we compare the Dyson-Maleev\cite{Dyson-Maleev} (DM)
and the Holstein-Primakoff\cite{HolsteinP40} (HP)
transformations as far as the ordering problem is concerned.
The DM transformation for the spin operators
$\hat{S}^{\pm}\equiv\hat{S}^x{\pm}i\hat{S}^y$ and $\hat{S}^z$,
in terms of bosonic operators $(\hat{a}^\dagger,\hat{a})$ is
\begin{eqnarray}
 \hat S^+&=&(2S)^{1/2} ~\hat a~,
 \nonumber \\
 \hat S^-&=&(2S)^{-{1\over2}} ~ \hat a^\dagger ~(2S-\hat a^\dagger\hat a) ~,
 \nonumber \\
 \hat S^z &=& S - \hat a^\dagger \hat a ~.
\label{e.DM}
\end{eqnarray}
The transformation is canonical, as from $[\hat{a},\hat{a}^\dagger]=1$
the spin commutation relations follow, with
$|\hat{\mbox{\boldmath$S$}}|^2=S(S+1)$.
Their Weyl symbols are found to be \cite{CTVV96prl}
\begin{eqnarray}
 S^+ &=& (2S)^{1/2} ~a ~,
\nonumber \\
 S^-&=&(2S)^{-{1\over2}} ~(2\widetilde S-a^*a)~a^* ~,
\nonumber \\
 S^z &=& \widetilde{S} - a^*a ~,
\label{e.DMWeyl}
\end{eqnarray}
For the use of the DM transformation in the 2DQHAF context we refer to Ref.
\onlinecite{CTVV96prl}.

The HP transformation reads instead
\begin{eqnarray}
 \hat S^+ &=& (2S-\hat a^\dagger \hat a)^{1/2}~\hat a ~,
 \nonumber \\
 \hat S^- &=& (\hat S^+)^\dagger~,
 \nonumber \\
 \hat S^z &=& S - \hat a^\dagger \hat a ~,
\label{e.HPada}
\end{eqnarray}
and can be rewritten in terms of phase-space operators
$\hat{q}=(\hat{a}^\dagger{+}\hat{a})/(2\widetilde{S})^{1/2}$ and
$\hat{p}=i(\hat{a}^\dagger{-}\hat{a})/(2\widetilde{S})^{1/2}$,
where $\widetilde{S}=S{+}1/2$ and with commutator
$[\hat{q},\hat{p}]=i/\widetilde{S}$, as
\begin{eqnarray}
 \hat{S}^+ &=& \widetilde{S} \bigg[1-{1\over 4}\bigg(
 \hat{q}^2+\hat{p}^2+{1\over\widetilde{S}}\bigg)\bigg]^{1/2}
 \left(\hat{q}+i\hat{p}\right)~,
\nonumber\\
 \hat{S}^-&=&(\hat{S}^+)^\dagger~,
\nonumber\\
\hat{S}^z&=&\widetilde{S}\left(1-{\hat{q}^2+\hat{p}^2 \over2}\right)~.
\label{e.HP}
\end{eqnarray}
We have now to determine the Weyl symbols of these operators;
indicating with $A_{\rm W}$ the Weyl symbol of the
operator $\hat{A}$, the following product relation\cite{Berezin80} holds:
\begin{equation}
 (AB)_{\rm W} = A_{\rm W} \exp\left[-{i\over2\widetilde{S}}
 ( \loarrow\partial_p \roarrow\partial_q
 - \loarrow\partial_q \roarrow\partial_p )\right]
 B_{\rm W}~.
\label{e.prodWeyl}
\end{equation}
Let us start with the operator $\hat{S}^+$, which has indeed the form of
the product of two Weyl-ordered operators,
$\hat{S}^+=\widetilde{S}\hat{A}\hat{B}$, with
\begin{equation}
 A_{\rm{W}} =
 \bigg[1-{1\over 4}\bigg(q^2+p^2+{1\over\widetilde{S}}\bigg)\bigg]^{1/2}
 \equiv \sum_n c_n\, (q^2+p^2)^n
\end{equation}
and $B_{\rm{W}}=q+i\,p$~, where the coefficients $c_n$ arise from the
expansion of the square root.
By using Eq.~(\ref{e.prodWeyl}) we then find
\begin{eqnarray}
 (AB)_{\rm{W}}
 &=& \sum_n c_n\,(q^2+p^2)^n
 e^{ -i (\loarrow\partial_p \roarrow\partial_q
 {-}\loarrow\partial_q \roarrow\partial_p)
 /2\widetilde{S} }\, (q{+}ip)
\nonumber\\
 &=& \sum_n c_n\,(q^2+p^2)^n
 \bigg[ q +ip - {1\over2\widetilde{S}} \big(\stackrel{\leftarrow}\partial_q
 + i\stackrel{\leftarrow}\partial_p \big) \bigg]
\nonumber\\
 &=& \sum_n c_n\,\bigg[ (q^2+p^2)^n - {n\over\widetilde{S}}
 (q^2+p^2)^{n{-}1} \bigg] \, (q{+}ip)
\nonumber\\
 &\approx& \sum_n c_n\,\bigg(q^2+p^2-{1\over\widetilde{S}}\bigg)^n \,(q{+}ip)
\nonumber\\
 &=& \bigg[1-{1\over4}(q^2+p^2)\bigg]^{1/2} \,(q{+}ip)~,
\label{e.S+Weyl1}
\end{eqnarray}
where terms up to the first order in $1/\widetilde{S}$ have been kept;
hence, being $S^-=(S^+)^*$ and the operator $\hat{S}^z$
in Eq.~(\ref{e.HP}) already Weyl
ordered, the Weyl symbols of the HP operators are
\begin{eqnarray}
 S^\pm &=&
 \widetilde{S}\,\bigg[1-{1\over4}(q^2+p^2)\bigg]^{1/2} \,(q\pm ip)~,
\nonumber\\
 S^z &=& \widetilde{S}\left(1-{q^2+p^2\over2}\right)~.
\label{e.HPWeyl}
\end{eqnarray}
Note that the value $\widetilde{S}\equiv{S+{1\over2}}$ appears as
the natural spin length of the theory, being
$|{\mbox{\boldmath$S$}}|^2=(S^z)^2+S^+S^-=\widetilde{S}^2$~, both in the DM
case (\ref{e.DMWeyl}) and in the HP case (\ref{e.HPWeyl}).

In the above derivation for the HP case we have neglected terms of the
order $1/\widetilde{S}^2$ and higher: this approximation is necessary to
obtain reasonably simple expressions of $S^\pm_{\rm W}$. Although no such
approximation seems to appear when using the DM transformation, this is in
fact a wrong conclusion.
Indeed, it is well known that, in order to apply the DM transformation, and
properly take into account the kinematic interaction, one should use,
rather than the simple transformed Hamiltonian $\hat{{\cal{H}}}_{\rm DM}$,
the operator $\hat{P}\hat{\cal{H}}_{\rm DM}\hat{P}$ where $\hat{P}$ is the
projector on the Hilbert subspace of the spin system. Such subspace is
generated by the eigenstates of the operator $\hat{a}^\dagger\hat{a}$ with
positive eigenvalues $n{\le}2S$, i.e., by the eigenstates of the operator
$\hat{z}^2\equiv(\hat{q}^2+\hat{p}^2)/2$ with $2\widetilde{S}$ equispaced
positive eigenvalues $0\le{z}^2\le(2-1/2\widetilde{S})$. On the other hand,
to determine the explicit form of the Weyl symbol for the operator
$\hat{P}\hat{\cal{H}}_{\rm DM}\hat{P}$ is an impossible task, unless one
only keeps terms up to the first order in $1/\widetilde{S}$; this means to
approximate $\hat{P}$ with the identity operator, which is in fact what we
have done in Sec.~\ref{s.eh}.
Both transformations can hence be used in the framework of the PQSCHA
and do actually involve the same semiclassical approximation.

\section{The effective spin Hamiltonian}
\label{a.esh}

The general expression given by the PQSCHA for the LCA effective
Hamiltonian is\cite{CTVV92ham}
\begin{equation}
 {\cal{H}}_{\rm eff}=
 e^\Delta{\cal{H}}({\mbox{\boldmath$p$}},{\mbox{\boldmath$q$}})
 -\sum_{\bf{k}}\alpha_{\bf{k}}\omega^2_{\bf{k}}
 +{1\over\beta}\sum_{\bf{k}}\ln{\sinh f_{\bf{k}}\over f_{\bf{k}}}~,
\label{e.Heffgen}
\end{equation}
where $\cal{H}({\mbox{\boldmath$p$}},{\mbox{\boldmath$q$}})$ is the Weyl
symbol of the original Hamiltonian $\hat{\cal{H}}$ and
\begin{eqnarray*}
 \Delta&=&{1\over2} \sum_{{\bf{i}}{\bf{j}}}\left[
 D^{(pp)}_{{\bf{i}}{\bf{j}}}\partial_{p_{\bf{i}}}\partial_{p_{\bf{j}}}+
 D^{(qq)}_{{\bf{i}}{\bf{j}}}\partial_{q_{\bf{i}}}\partial_{q_{\bf{j}}}
 \right]~,
\\
 D^{(pp)}_{{\bf{i}}{\bf{j}}}&=& {1\over N}\sum_{\bf{k}}
 b^2_{\bf{k}}\,\alpha_{\bf{k}}\,\cos{\bf{k}}{\cdot}({\bf{i}}{-}{\bf{j}})~,
\\
 D^{(qq)}_{{\bf{i}}{\bf{j}}}&=& {1\over N}\sum_{\bf{k}}
 a^2_{\bf{k}}\,\alpha_{\bf{k}}\,\cos{\bf{k}}{\cdot}({\bf{i}}{-}{\bf{j}})~,
\\
 \alpha_{\bf{k}}&=&{\hbar\over2\omega_{\bf{k}}}{\cal{L}}_{\bf{k}}~,
\\
 {\cal{L}}_{\bf{k}}&=&\coth f_{\bf{k}} -{1\over f_{\bf{k}}}~,
\\
 f_{\bf{k}}&=&{\beta\hbar\omega_{\bf{k}} \over2}~,
\\
 \omega_{\bf{k}}&=&a_{\bf{k}} b_{\bf{k}}~,
\\
 a^2_{\bf{k}}&=&{1\over N}\sum_{{\bf{i}}{\bf{j}}}e^{i{\bf{k}}{\cdot}
 ({\bf{i}}-{\bf{j}})}A^2_{{\bf{i}}{\bf{j}}}~,
\\
 b^2_{\bf{k}}&=&{1\over N}\sum_{{\bf{i}}{\bf{j}}}e^{i{\bf{k}}{\cdot}
 ({\bf{i}}-{\bf{j}})}B^2_{{\bf{i}}{\bf{j}}}~,
\end{eqnarray*}
where ${\bf{i}}$ and ${\bf{j}}$ are sites on the lattice, while
$A^2_{{\bf{i}}{\bf{j}}}$ and $B^2_{{\bf{i}}{\bf{j}}}$ are the LCA
approximations of the fundamental renormalization parameters of the theory:
\begin{eqnarray*}
 A^2_{{\bf{i}}{\bf{j}}}({\mbox{\boldmath$p$}},{\mbox{\boldmath$q$}})
 \equiv\partial_{p_{\bf{i}}}\partial_{p_{\bf{j}}}
 e^\Delta{\cal{H}}({\mbox{\boldmath$p$}},{\mbox{\boldmath$q$}})
 \simeq A^2_{{\bf{i}}{\bf{j}}}~,
\\
 B^2_{{\bf{i}}{\bf{j}}}({\mbox{\boldmath$p$}},{\mbox{\boldmath$q$}})
 \equiv \partial_{q_{\bf{i}}}\partial_{q_{\bf{j}}}
 e^\Delta{\cal{H}}({\mbox{\boldmath$p$}},{\mbox{\boldmath$q$}})
 \simeq B^2_{{\bf{i}}{\bf{j}}}~;
\end{eqnarray*}
their dependence upon the phase-space coordinate
$({\mbox{\boldmath$p$}},{\mbox{\boldmath$q$}})$ is eliminated by the LCA.

As suggested in Ref.~\onlinecite{CTVV92ham}, and for the reasons given in
Sec.~\ref{s.eh}, in this work we define the specific LCA to be used, by
setting
$A^2_{{\bf{i}}{\bf{j}}}({\mbox{\boldmath$p$}},{\mbox{\boldmath$q$}})\approx
\langle{A}^2_{{\bf{i}}{\bf{j}}}({\mbox{\boldmath$p$}},{\mbox{\boldmath$q$}})
\rangle^{\rm{SCHA}}_{\rm{eff}}$ [and similarly for
$B^2_{{\bf{i}}{\bf{j}}}({\mbox{\boldmath$p$}},{\mbox{\boldmath$q$}})$],
where $\langle\cdots\rangle^{\rm SCHA}_{\rm eff}$ is the SCHA approximation
of the classical-like average  $\langle\cdots\rangle_{\rm eff}$ with the
effective Hamiltonian, Eq.~(\ref{e.aveOpq}) below. We then have (see
Ref.~\onlinecite{CTVV92ham})
\begin{eqnarray*}
 A^2_{{\bf{i}}{\bf{j}}}&=&
 \big\langle\partial_{p_{\bf{i}}} \partial_{p_{\bf{j}}}
 e^\Delta{\cal{H}}({\mbox{\boldmath$p$}},{\mbox{\boldmath$q$}})
 \big\rangle^{\rm SCHA}_{\rm eff}~,
\\
 B^2_{{\bf{i}}{\bf{j}}}&=&
 \big\langle\partial_{q_{\bf{i}}} \partial_{q_{\bf{j}}}
 e^\Delta{\cal{H}}({\mbox{\boldmath$p$}},{\mbox{\boldmath$q$}})
 \big\rangle^{\rm SCHA}_{\rm eff}~~.
\end{eqnarray*}

Let us consider now the case of the spin Hamiltonian~(\ref{e.ham}).
We consider the lattice as subdivided into the usual AFM positive and
negative sublattices and hereafter use the notation $(-)^{\bf{i}}={\pm}1$
for the site ${\bf{i}}$ belonging to the former or the latter, respectively.
The Weyl symbol for $\hat{\cal{H}}$ (using the DM transformation) is given
by \cite{CTVV96prl}
\begin{eqnarray}
 {{\cal H}\over J\widetilde{S}^2} &=& -{1\over2} \sum_{{\bf{i}},{\bf{d}}}
 \bigg[ (1-z_{\bf{i}}^2)(1-z_{{\bf{i}}+{\bf{d}}}^2)
\nonumber\\
 & &+\bigg(1-{z_{\bf{i}}^2+z_{{\bf{i}}+{\bf{d}}}^2\over4}\bigg)
 ~(q_{\bf{i}}q_{{\bf{i}}+{\bf{d}}}-p_{\bf{i}}p_{{\bf{i}}+{\bf{d}}})
\nonumber\\
 & &+i~(-)^{\bf{i}}{z_{\bf{i}}^2-z_{{\bf{i}}+{\bf{d}}}^2\over4}
 ~(q_{\bf{i}}p_{{\bf{i}}+{\bf{d}}}+p_{\bf{i}}q_{{\bf{i}}+{\bf{d}}}) \bigg]~,
\label{e.ham.Weyl}
\end{eqnarray}
where $(p_{\bf{i}},q_{\bf{i}})$ are the
Weyl symbols of canonical operators $(\hat{p}_{\bf{i}},\hat{q}_{\bf{i}})$
such that $[\hat{q}_{\bf{i}},\hat{p}_{\bf{j}}]=
i\widetilde{S}^{-1}\delta_{\bf{ij}}$, and
$z_{\bf{i}}^2\equiv(q_{\bf{i}}^2+p_{\bf{i}}^2)/2$.
It is simpler to use $J\widetilde{S}^2$ as the energy unit, i.e., to apply
the above framework to the dimensionless Hamiltonian
${\cal{H}}({\mbox{\boldmath$p$}},{\mbox{\boldmath$q$}})/J\widetilde{S}^2$,
so that all the relevant quantities are dimensionless; in particular,
$\beta\to1/t$ and $\hbar\to1/\widetilde{S}$.

The first step to close the self-consistent scheme described above
is then the evaluation of
$e^\Delta\cal{H}({\mbox{\boldmath$p$}},{\mbox{\boldmath$q$}})$.
By performing the transformation ${\bf{k}}\to(\pi,\pi)-{\bf{k}}$ one can
establish the identity
\begin{equation}
 D^{(qq)}_{{\bf{i}},{\bf{i}}+{\bf{r}}}
 = (-)^{r_1{+}r_2}D^{(pp)}_{{\bf{i}},{\bf{i}}+{\bf{r}}}~,
\end{equation}
where ${\bf{r}}=(r_1,r_2)$; in particular, only the following two
renormalization parameters appear in the effective Hamiltonian:
\begin{eqnarray}
 D &\equiv& D^{(qq)}_{{\bf{i}}{\bf{i}}}
 = D^{(pp)}_{{\bf{i}}{\bf{i}}}~,
\nonumber\\
 D' &\equiv& D^{(qq)}_{{\bf{i}},{\bf{i}}+{\bf{d}}}
 = - D^{(pp)}_{{\bf{i}},{\bf{i}}+{\bf{d}}}~,
\end{eqnarray}
where ${\bf{d}}=(\pm{1},0)$ or $(0,\pm{1})$ is a nearest-neighbour
displacement. We then find, for nearest neighbours
${\bf{i}}$ and ${\bf{j}}={\bf{i}}+{\bf{d}}$,
\begin{eqnarray*}
 e^\Delta \,{z}^2_{\bf{i}} &=& z^2_{\bf{i}} +D~,
\\
 e^\Delta \,{q}_{\bf{i}}q_{\bf{j}} &=& q_{\bf{i}}q_{\bf{j}} +D'~,
\\
 e^\Delta \,{p}_{\bf{i}}p_{\bf{j}} &=& p_{\bf{i}}p_{\bf{j}} -D'~;
\end{eqnarray*}
furthermore,
\[
 e^\Delta z^2_{\bf{i}} z^2_{\bf{j}} =
 z^2_{\bf{i}} z^2_{\bf{j}} {+} D (z^2_{\bf{i}} {+} z^2_{\bf{j}})
 + D'(q_{\bf{i}} q_{\bf{j}} {-} p_{\bf{i}} p_{\bf{j}})
 {+}D^2{+}D'^2
\]
and
\begin{eqnarray*}
 e^\Delta (z^2_{\bf{i}}{+}z^2_{\bf{j}})
 (q_{\bf{i}} q_{\bf{j}} {-} p_{\bf{i}} p_{\bf{j}}) &=&
 (z^2_{\bf{i}}{+}z^2_{\bf{j}}{+}4D)
 (q_{\bf{i}} q_{\bf{j}} {-} p_{\bf{i}} p_{\bf{j}})
\\
  & & \hskip 5truemm
 + 4D'(z^2_{\bf{i}}{+}z^2_{\bf{j}}) {+} 8DD'~,
\\
 e^\Delta (z^2_{\bf{i}}{-}z^2_{\bf{j}})
 (q_{\bf{i}} q_{\bf{j}} {+} p_{\bf{i}} p_{\bf{j}}) &=&
 (z^2_{\bf{i}}{-}z^2_{\bf{j}})
 (q_{\bf{i}} q_{\bf{j}} {+} p_{\bf{i}} p_{\bf{j}})~.
\end{eqnarray*}
It is important that the imaginary part of
${\cal{H}}({\mbox{\boldmath$p$}},{\mbox{\boldmath$q$}})$ does not
contribute any renormalization term: this in fact assures the final
effective spin Hamiltonian, i.e., the one obtained after having performed
the classical version of the inverse of the DM transformation on
$\cal{H}_{\rm{eff}}({\mbox{\boldmath$p$}},{\mbox{\boldmath$q$}})$, to be
real.
By defining $\theta^2\equiv{1}{-}{\cal{D}}/2$ with ${\cal{D}}\equiv(D{-}D')$,
we find
\begin{eqnarray*}
 { e^\Delta{\cal H}({\mbox{\boldmath$p$}},{\mbox{\boldmath$q$}})
 \over J\widetilde{S}^2 } &=&
 -{1\over2} \sum_{{\bf{i}},{\bf{d}}}
 \bigg[\theta^4-\theta^2(z_{\bf{i}}^2+z_{{\bf{i}}+{\bf{d}}}^2) +
 z_{\bf{i}}^2z_{{\bf{i}}+{\bf{d}}}^2
\\
 & & \hskip 7truemm
 +\bigg(\theta^2-{z_{\bf{i}}^2+z_{{\bf{i}}+{\bf{d}}}^2\over4}\bigg)
 ~(q_{\bf{i}}q_{{\bf{i}}+{\bf{d}}}-p_{\bf{i}}p_{{\bf{i}}+{\bf{d}}})
\\
 & & \hskip 7truemm
 +i~(-)^{\bf{i}}{z_{\bf{i}}^2-z_{{\bf{i}}+{\bf{d}}}^2\over4}
 ~(q_{\bf{i}}p_{{\bf{i}}+{\bf{d}}}+p_{\bf{i}}q_{{\bf{i}}+{\bf{d}}}) \bigg]~;
\end{eqnarray*}
this is the only term of
${\cal{H}}_{\rm{eff}}({\mbox{\boldmath$p$}},{\mbox{\boldmath$q$}})$ we need
to determine $a^2_{\bf{k}}$ and $b^2_{\bf{k}}$ (as
configuration-independent terms do not enter the evaluation of
$\langle\cdots\rangle^{\rm SCHA}_{\rm eff}$), which are easily found to be
\begin{eqnarray}
 a^2_{\bf{k}}&=&4\kappa^2(1+\gamma_{\bf{k}})
\nonumber\\
 b^2_{\bf{k}}&=&4\kappa^2(1-\gamma_{\bf{k}})
\end{eqnarray}
where $\gamma_{\bf{k}}=(\cos{k_1}+\cos{k_2})/2$ and we introduced the SCHA
renormalization parameter
\begin{equation}
 \kappa^2= 1-{1\over2N\widetilde{S}}
 \sum_{\bf{k}}(1-\gamma^2_{\bf{k}})^{1/2}\coth f_{\bf{k}}~,
\label{e.kappa}
\end{equation}
that can also be written as in Eq.~(\ref{e.kappasc}); the renormalized
frequencies $\omega_{\bf{k}}$ are hence given by Eq.~(\ref{e.omegasc}); the
dimensionless parameter $f_{\bf{k}}$ can be written as
$f_{\bf{k}}=\omega_{\bf{k}}/(2\widetilde{S}t)$. With the above
determinations one can express $\cal{D}$ as in Eq.~(\ref{e.D}), and the
procedure is completed by the self-consistent solution of
Eqs.~(\ref{e.kappasc}) and~(\ref{e.omegasc}). As for the first uniform term
appearing in Eq.~(\ref{e.Heffgen}), one easily finds
\begin{equation}
 {1\over N} \sum_{\bf{k}} \omega^2_{\bf{k}} \alpha_{\bf{k}}
 =  {t\over N} \sum_{\bf{k}} f_{\bf{k}} {\cal{L}}_{\bf{k}}
 = 2 \kappa^2 {\cal{D}}~.
\end{equation}
To recast $e^\Delta{\cal{H}}({\mbox{\boldmath$p$}},{\mbox{\boldmath$q$}})$
in the form of a spin Hamiltonian, we scale
$({\mbox{\boldmath$p$}},{\mbox{\boldmath$q$}})\rightarrow
(\theta{\mbox{\boldmath$p$}},\theta{\mbox{\boldmath$q$}})$
so that
\begin{eqnarray*}
 { e^\Delta{\cal H}({\mbox{\boldmath$p$}},{\mbox{\boldmath$q$}})
 \over J\widetilde{S}^2 } &=&
 {\theta^4\over2}\sum_{{\bf{i}},{\bf{d}}}
 \bigg[(1-z_{\bf{i}}^2)(1-z_{{\bf{i}}+{\bf{d}}}^2)
\\
 & & \hskip 7truemm
 +\bigg(1-{z_{\bf{i}}^2+z_{{\bf{i}}+{\bf{d}}}^2\over4}\bigg)
 ~(q_{\bf{i}}q_{{\bf{i}}+{\bf{d}}}-p_{\bf{i}}p_{{\bf{i}}+{\bf{d}}})
\\
 & & \hskip 7truemm
 +i~(-)^{\bf{i}}{z_{\bf{i}}^2-z_{{\bf{i}}+{\bf{d}}}^2\over4}
 ~(q_{\bf{i}}p_{{\bf{i}}+{\bf{d}}}+p_{\bf{i}}q_{{\bf{i}}+{\bf{d}}})
\bigg]~;
\end{eqnarray*}
this equation has the same functional form of the Weyl symbol for the
Hamiltonian, Eq.~(\ref{e.ham.Weyl}), so that performing the inverse of the
classical DM transformation we eventually find the form reported in
Eq.~(\ref{e.Heff}), where the change in the phase-space measure by the
factor $\theta^{2N}$ due to the above scaling is absorbed in the effective
Hamiltonian as an additive logarithmic term, $-Nt\ln\theta^2$.

Finally, the PQSCHA expresses the general thermal average of an observable
$\hat{\cal{O}}(\hat{\mbox{\boldmath$p$}},\hat{\mbox{\boldmath$q$}})$ as the
classical-like average with the effective Hamiltonian
\begin{equation}
 \langle\hat {\cal O}\rangle={1\over {\cal Z}}
 \int {d{\mbox{\boldmath$p$}}\,d{\mbox{\boldmath$q$}} \over(2\pi)^N}
 ~\widetilde{\cal{O}}({\mbox{\boldmath$p$}},{\mbox{\boldmath$q$}})
 ~e^{\displaystyle -\beta {\cal H}_{\rm{eff}}({\mbox{\boldmath$p$}},{\mbox{\boldmath$q$}})}
 \equiv\langle\widetilde{\cal{O}}\rangle_{\rm{eff}}~,
\label{e.aveOpq}
\end{equation}
of the phase-space function
\begin{equation}
 \widetilde{\cal{O}}({\mbox{\boldmath$p$}},{\mbox{\boldmath$q$}})\equiv
 e^\Delta{\cal{O}}({\mbox{\boldmath$p$}},{\mbox{\boldmath$q$}})~,
\end{equation}
where ${\cal O}({\mbox{\boldmath$p$}},{\mbox{\boldmath$q$}})$ is the Weyl
symbol of the operator $\hat{\cal O}$. In terms of the classical spin
variables Eq.~(\ref{e.aveOpq}) becomes just Eq.~(\ref{e.aveO}).
In the case of the correlation functions
$G({\bf{r}})=\langle\hat{\mbox{\boldmath$S$}}_{\bf{i}}{\cdot}
\hat{\mbox{\boldmath$S$}}_{{\bf{i}}+{\bf{r}}}\rangle$
the procedure leading to the Weyl symbol
$S_{{\bf{i}},{\bf{i}}+{\bf{r}}}({\mbox{\boldmath$p$}},{\mbox{\boldmath$q$}})$
of the operator $\hat{\mbox{\boldmath$S$}}_{\bf{i}}{\cdot}
\hat{\mbox{\boldmath$S$}}_{{\bf{i}}+{\bf{r}}}$
and hence to the expression of $e^\Delta
S_{{\bf{i}},{\bf{i}}+{\bf{r}}}({\mbox{\boldmath$p$}},{\mbox{\boldmath$q$}})$
is obviously analogous to the one described for the Hamiltonian, and the
final result is easily found to be as reported in Eq.~(\ref{e.Gdr}).


\end{document}